\documentclass[aps,pra,notitlepage,superscriptaddress,10pt]{revtex4-1}
\usepackage[utf8]{inputenc}
\usepackage{graphicx}		
\usepackage{bm} 		
\usepackage{caption}
\usepackage{xcolor}
\usepackage{siunitx}

\begin{document}

\title{Low-noise heralded single photons from cascaded downconversion}
\author{Patrick F. Poitras}
\affiliation{Département de Physique et d’Astronomie, Université de Moncton, 18 Ave. Antonine-Maillet, Moncton, New Brunswick E1A 3E9, Canada}
\author{Evan Meyer-Scott}
\affiliation{Integrated Quantum Optics, Department of Physics, University of Paderborn, Warburger Strasse 100, 33098 Paderborn, Germany}
\author{Deny R. Hamel}
\email{Corresponding author: deny.hamel@umoncton.ca}
\affiliation{Département de Physique et d’Astronomie, Université de Moncton, 18 Ave. Antonine-Maillet, Moncton, New Brunswick E1A 3E9, Canada}

\begin{abstract}
Heralded single photon sources are often implemented using spontaneous parametric downconversion, but their quality can be restricted by optical loss, double pair emission and detector dark counts. Here, we show that the performance of such sources can be improved using cascaded downconversion, by providing a second trigger signal to herald the presence of a single photon, thereby reducing the effects of detector dark counts. We find that for a setup with fixed detectors, an improved heralded second-order correlation function $g^{(2)}$ can always be achieved with cascaded downconversion given sufficient efficiency for the second downconversion, even for equal single-photon production rates. Furthermore, the minimal $g^{(2)}$ value is unchanged for a large range in pump beam intensity. These results are interesting for applications where achieving low, stable values of $g^{(2)}$ is of primary importance.
\end{abstract}

\maketitle

\section{Introduction}

\begin{figure}[ht!]
\centering \includegraphics[width=\textwidth]{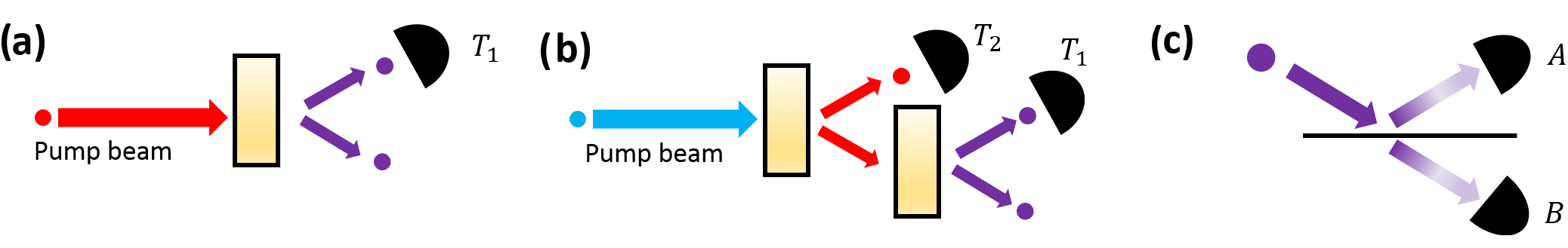}
\caption[]{Scheme to produce heralded single photons using (a) a normal downconversion source and a heralding detector, $T_1$ and (b) cascaded downconversion, where the output of the first downconversion acts as the pump for the second, heralded with heralding detectors $T_1$ and $T_2$.   
(c)~Scheme for the evaluation of the single photon character of the source. The produced heralded photons are sent on a 50:50 beamsplitter with $g^{(2)}$ detectors $A$ and $B$ on each end.}\noindent \label{fig:scheme}
\end{figure}

Single photons constitute an important resource for several quantum optical applications, such as optical quantum computing, randomness generation or quantum communication \cite{Eisaman2011}.
An ideal source would be deterministic, producing single photons on demand. However, heralded single photons sources, which produce single photons at random times accompanied by a heralding signal that announces the photon's creation, are sufficient for many applications\cite{Eisaman2011,Schiavon2016}. A common way to implement heralded sources is through the process of spontaneous parametric downconversion (SPDC). As shown in Figure~\ref{fig:scheme}~(a), SPDC produces photons in pairs, thus the detection of one photon announces the presence of the other.

However, the quality of heralded single photons produced by SPDC is limited, since a detection at the heralding detector is not always associated with a single photon. Indeed, there could be no photon present, due to optical losses and detector dark counts, or more than one photon present because of double-pair emission \cite{Migdall2002a}. 
Moreover, for applications where pure single photons are required, the spectral correlations resulting from SPDC can also be detrimental\cite{Grice2001,URen2005,Mosley2008a,Mosley2008,Levine2010}.

Many strategies have been proposed to improve the quality of SPDC single photon sources. These include for example various multiplexing schemes, which aim to parallelize single photon production using various degrees of freedom \cite{Migdall2002a,Jeffrey2004,Shapiro2007,Ma2011,Collins2013,Xiong2016,Mendoza2016,Puigibert2017,Kaneda2015,Schiavon2016}, and the production of spectrally pure states by controlling the modal structure of the photon pair emission  \cite{URen2005,Mosley2008a,Mosley2008,Levine2010}.

Here, we investigate a different approach to improve the statistics of heralded single photons from SPDC, which is to use photon precertification \cite{Cabello2012}, implemented with cascaded downconversion (CSPDC) as shown in Figure~\ref{fig:scheme}~(b). CSPDC has already been demonstrated using separate nonlinear waveguides \cite{Hubel2010}, from an integrated on-chip setup \cite{Krapick2016} and with a hybrid system using rubidium vapor and a $\chi^{(2)}$ nonlinear waveguide \cite{Ding2015}, and has also been used for photon precertification \cite{Meyer-Scott2016}. In the context of single-photon heralding, CSPDC is equivalent to pumping a regular SPDC source with heralded single photons, resulting in photon pairs with fundamentally different statistics. At low photon rates, the main benefit is to provide a second heralding signal which can reduce the negative effects of dark counts at the heralding detectors. This approach can potentially result in single photons with a lower heralded second-order correlation function $g^{(2)}$\cite{Grangier1986}, given by

\begin{equation}\label{eq:g2}
g^{(2)} = \frac{T_\mathrm{1AB} S_\mathrm{1}}{D_{\mathrm{1A}} D_{\mathrm{1B}}},
\end{equation}

\noindent where $T_\mathrm{1AB}$ represents the rate of triplets, $S_\mathrm{1}$ the rate of singles and where $D_{\mathrm{1A}}$ and $D_{\mathrm{1B}}$ represents pair production rates. Assuming no additional noise, the heralded $g^{(2)}$ quantifies the pollution of the single photon state by unwanted events such as multiple photons from higher-order terms in the SPDC process, as well as false heralding, where the presence of a non-existing photon is heralded, and accidental coincidences, when SPDC pairs are produced near-simultaneously by coincidence. A value of $g^{(2)} = 0$ represents a perfect single photon uncontaminated with additional photons, a characteristic which is important in applications such as quantum key distribution where multi-photon events can reduce the security of the system\cite{Brassard2000,Schiavon2016}, or in linear optical quantum computing, where operating outside the post-selection basis requires a low $g^{(2)}$\cite{Jennewein2011}. Experimentally, it has been shown to be possible to reduce the $g^{(2)}$ down to $7\mathrm{x}10^{-4}$ for heralded single photon sources \cite{Eisaman2011} and down to less than $3\mathrm{x}10^{-3}$ for semiconductor quantum-dot single photon sources\cite{Senellart2017}.

Here we examine the possible advantages of cascaded downconversion for heralding single photons. In other words, given a fixed SPDC-based heralded single photon source and detectors, is it advantageous to pump it using heralded single photons rather than just using a coherent beam from a laser? We address these questions first using an analytic treatment in section~\ref{sec:analytical}, followed by a more exact numerical simulation in section~\ref{sec:simulations} based on the detector statistic model introduced by Bussières et al.\cite{Bussieres2008}.

\section{Analytical Model}\label{sec:analytical}

We can use an approximate analytic treatment to estimate each term in equation~\ref{eq:g2} and thus determine situations where CSPDC can be advantageous. 
We consider a heralded single photon produced by regular SPDC, as shown in Figure~\ref{fig:scheme}~(a). The singles rate at the heralding detector, $S_\mathrm{1}$, and at the $g^{(2)}$ detectors, $S_\mathrm{A}$ and $S_\mathrm{B}$, are given by

\begin{eqnarray}\label{eq:singles}
S_\mathrm{1} &=& N \eta_1 +d_1   \\
S_\mathrm{A} = S_\mathrm{B} &=& \frac{N \eta_{AB} }{2} +d_{AB},
\end{eqnarray}

\noindent where $\eta_i$ is the Klyshko effiency\cite{Klyshko1980} and $d_i$ is the detector dark count rate for detector $i$, assumed to be the same for the $A$ and $B$ detectors, and $N$ is the rate of photon pairs produced in the crystal. Since we seek a situation where $N$ is low in order to produce a low $g^{(2)}$, it is safe to assume that accidental coincidences have a negligible impact on the two-fold coincidence rate. The rate of these two-fold coincidences will be
\begin{equation}\label{eq:coinc}
D_{\mathrm{1A}} = D_{\mathrm{1B}} = \frac{N \eta_1 \eta_{AB}}{2}.
\end{equation}

\noindent Finally, the rate of three-fold coincidences are given by
\begin{equation}\label{eq:triples}
T_{\mathrm{1AB}} = D_{\mathrm{1A}} S_\mathrm{B} W + D_{\mathrm{1B}} S_\mathrm{A} W - D_{\mathrm{1A}} D_{\mathrm{1B}} W ,
\end{equation}

\noindent where $W$ is the coincidence window of the detectors. The last term in equation~\ref{eq:triples} ensures that double pairs leading to a three-fold coincidence are not counted twice. Substituting the equations~\ref{eq:singles},  \ref{eq:coinc} and \ref{eq:triples} into equation~\ref{eq:g2} gives the expected heralded second order correlation for SPDC:
\begin{equation}\label{eq:g2_spdc}
g^{(2)}_\mathrm{S} = \left( \frac{4 d_{AB}}{N \eta_1\eta_{AB}} + \frac{2}{\eta_1} -1 \right) \left(N W \eta_1 + W d_1\right).
\end{equation}

Since our goal is to optimize the $g^{(2)}_\mathrm{S}$ value that such a source can produce, we treat the rate of pair creation, which can easily be tuned by changing the intensity of the pump, as a free parameter. Finding the value of $N$ which minimizes $g^{(2)}_\mathrm{S}$ and substituting it in equation \ref{eq:g2_spdc} results in an optimal second-order correlation function of
\begin{equation}\label{eq:spdc_g2_min}
g^{(2)}_\mathrm{S,min} = \left(\frac{2}{\sqrt{H_{AB}}} + \sqrt{\frac{2-\eta_1}{H_1}} \right)^2 ,
\end{equation}

\noindent where $H=\frac{\eta}{W d}$ is the \textit{figure of merit} of the detectors, a property that depends on the detector model\cite{Hadfield2009,Eisaman2011}.

We now consider the second scenario, in which the same SPDC source is pumped with heralded single photons from another SPDC process, as shown in Figure~\ref{fig:scheme}~(b). In this case, equation~\ref{eq:g2} becomes 
\begin{equation}
g^{(2)}_\mathrm{C} = \frac{F D_{\mathrm{12}}}{T_{12A} T_{12B}},
\end{equation}

\noindent where $F$ is the fourfold rate.

The rate of detected triples (both heralding detectors and either $g^{(2)}$ detector) should be dominated by genuine triplets, so that the rates are given by
\begin{equation}\label{eq:cspdc-triples}
T_\mathrm{12A} = T_\mathrm{12B} = \frac{ N P \eta_1 \eta_2 \eta_{AB} }{2} ,
\end{equation}

\noindent where $P$ denotes the conversion efficiency of the second SPDC crystal. The rate of doubles, corresponding to a coincidence detection between both heralding detectors, is composed of pairs from a genuine cascaded downconversion event and from accidental coincidences on the heralding detectors. Since the SPDC processes are very inefficient, we expect the detection at every detector other than detector 2 to be dominated by dark counts, so that $S_{\mathrm{1}} \approx d_1.$ Detector 2, like detector 1 in the SPDC case, inherits the singles counts given by equation \ref{eq:singles}

\begin{equation}\label{eq:cspdc-triggerpairs}
D_\mathrm{12} = S_2 S_1 W + N P \eta_1 \eta_2 \approx S_2 d_1 W + N P \eta_1 \eta_2.
\end{equation}

\noindent Finally, most four-folds will be produced by an accidental coincidence between a real triplet and a dark count at either detector A or B or by a accidental coincidence due to the production of two pairs in the same coincidence window. As such, we have

\begin{equation}\label{eq:cspdc-fourfold}
F =  2 T_{\mathrm{12A}} d_{AB} W + \frac{(1-(1-\eta_2)^2)(1-(1-\eta_1)^2) N^2 \eta_{AB}^2 W}{2}.
\end{equation}

Substituting the values from equations~\ref{eq:cspdc-triples},  \ref{eq:cspdc-triggerpairs} and \ref{eq:cspdc-fourfold}, and performing the same optimization as above yields
\begin{equation}\label{eq:cspdc_g2_min}
g^{(2)}_{C,\mathrm{min}} =  \left(\frac{2}{\sqrt{H_{AB}}}\sqrt{1+\frac{1}{PH_1}}+\sqrt{\frac{(2-\eta_1)(2-\eta_2)}{H_1 H_2}}\right)^2.
\end{equation}

The second term in the previous equation comes from the interaction of two pairs in the same coincidence window, and is negligible unless the performance of the $g^{(2)}$ detectors is significantly better than that of the heralding detectors.

Neglecting the second term, equations~\ref{eq:spdc_g2_min} and \ref{eq:cspdc_g2_min} can be combined to show that the cascaded approach is advantageous if
\begin{equation}\label{eq:criterion}
P > \left(H_1 + \frac{2-\eta_1}{4} H_{AB} + \sqrt{(2-\eta_1)H_{AB}H_1}\right)^{-1}.
\end{equation}

\begin{figure}[h!]
\begin{minipage}{\linewidth}
\begin{center}
    \includegraphics[width=3.5in]{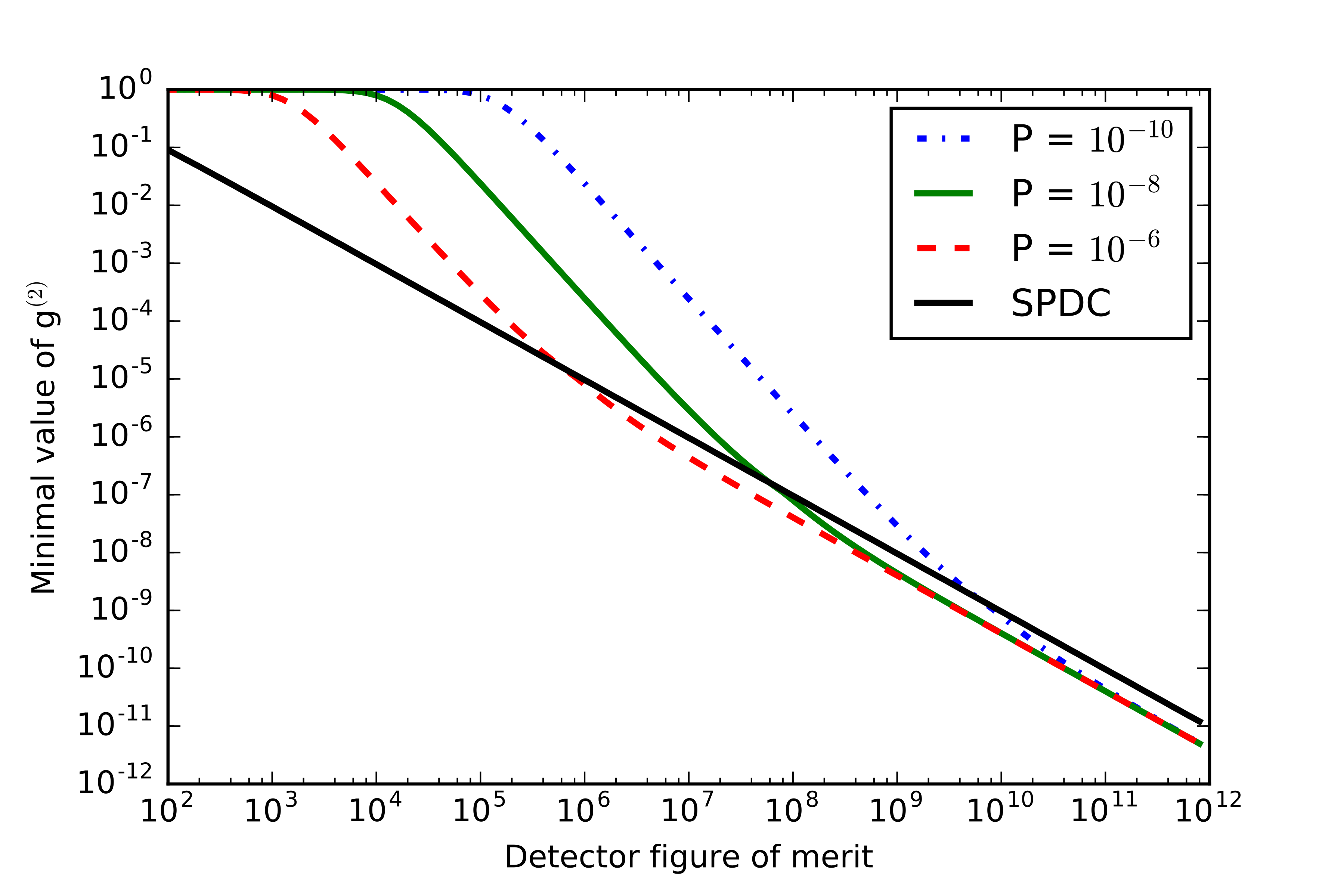}
    \captionof{figure}{\label{fig:H_g2_min} An example of the minimal $g^{(2)}$ for identical detectors. The necessity for high conversion efficiency in the SPDC crystal becomes less pronounced as the detector figure of merit increases. In this case, for superconducting nanowire detectors, which can have figures of merit above $10^{9}$\cite{Eisaman2011}, secondary downconversion requires an efficiency of $10^{-8}$ or better for CSPDC to be advantageous. This criterion is easily met by efficient SPDC crystals such as lithium niobate waveguides\cite{Hubel2010}.}
\end{center}
\end{minipage}
\end{figure}

Therefore, for given figures of merit $H_{AB}$ and $H_1$, an approach using cascaded downconversion will be advantageous as long as the probability of a secondary downconversion is sufficiently high, as shown in figure \ref{fig:H_g2_min}.

\section{Numerical modeling of detector behavior}\label{sec:simulations}

For a more exact modeling of detector behavior, we use an adapted version of the bucket-detector matrix formalism introduced in Bussières et. al. \cite{Bussieres2008} to accurately simulate the detector behavior.

This models detector statistics by taking a column vector of the form
\begin{equation}
\textbf{P} = [p_{\bar{A}\bar{B}\bar{1}},p_{\bar{A}\bar{B}1},...,p_{AB1}]^T,
\end{equation}

\noindent with the terms $p_{xyz}$ representing the probability of finding $g^{(2)}$ detectors A, B and heralding detector 1 in the states $x$,$y$,$z$. For example the $p_{\bar{A}B1}$ term represents the probability of a detection on detector B and the heralding detector, but not on detector A. This naturally extends to four detectors for CSPDC.

The simulation begins with a column vector that represents all detectors being in the off state

\noindent
\begin{equation}
\textbf{P}_0 = [1,0,0,...,0]^T.
\end{equation}

\noindent This vector can be modified by either photon detections or dark counts, respectively modeled by the detector transition matrices $M_\eta$ and $M_d$. If $i$ photon pairs are produced, the detector state becomes

\begin{equation}
\textbf{P'}_i = {M_\eta}^{i}M_d\textbf{P}_0.
\end{equation}

The final detector-state vector is given by a sum over all possible $P'_i$, weighed by $p_i$, the probability of obtaining $i$ photon pairs within a coincidence window. The final state is

\begin{equation}
\textbf{P} = \sum_{i=0}^{n}p_i\textbf{P'}_i = \sum_{i=0}^{n}p_i{M_\eta}^{i}M_d\textbf{P}_0.
\end{equation}

\noindent For our simulations, the $p_i$ follow the Poissonian statistics expected for distinguishable photon triplets\cite{Takesue2010}, reflecting the fact that until now experimental realizations of CPSDC have displayed high spectral correlations\cite{Shalm2013}. The case of indistinguishable single photons\cite{Zhang2018} can also be considered by changing $p_i$ accordingly, although at low photon pair rates this does not meaningfully impact results. While $n$ is infinite in the ideal case, it is here truncated when the probability of obtaining $n$ photons during a coincidence window becomes low enough as to not influence the detector statistics.  

\subsection{Modeling the $g^{(2)}$ of the output photons}

Considering that the $g^{(2)}$ detectors are only needed to measure the $g^{(2)}$, the properties of these detectors are not intrinsic to the source. As such, we can consider the output of the source using perfect detectors that do not have any dark counts.
In the limit where $d_{AB} \to 0$, $H_{AB} \to \infty$ and therefore equation \ref{eq:criterion} is trivially met if $P>0$.
In such a case, only multiple-photon events will contribute to the $T_{1AB}$ and $F$ terms for SPDC and CSPDC respectively. 

Analytically, equations \ref{eq:spdc_g2_min} and  \ref{eq:cspdc-fourfold} become
\begin{eqnarray}
g_{S,min}^{(2)} = \frac{2-\eta_1}{H_1}\\
g_{C,min}^{(2)} = \frac{(2-\eta_1)(2-\eta_2)}{H_1H_2}.
\end{eqnarray}

Therefore, the improvement in the $g^{(2)}$ for the CSPDC case is given by
\begin{equation}
\frac{g^{(2)}_{S,min}}{g^{(2)}_{C,min}} = \frac{H_2}{2-\eta_2}.
\end{equation}

\begin{figure}[h!]
\centering \includegraphics[width=3.5in]{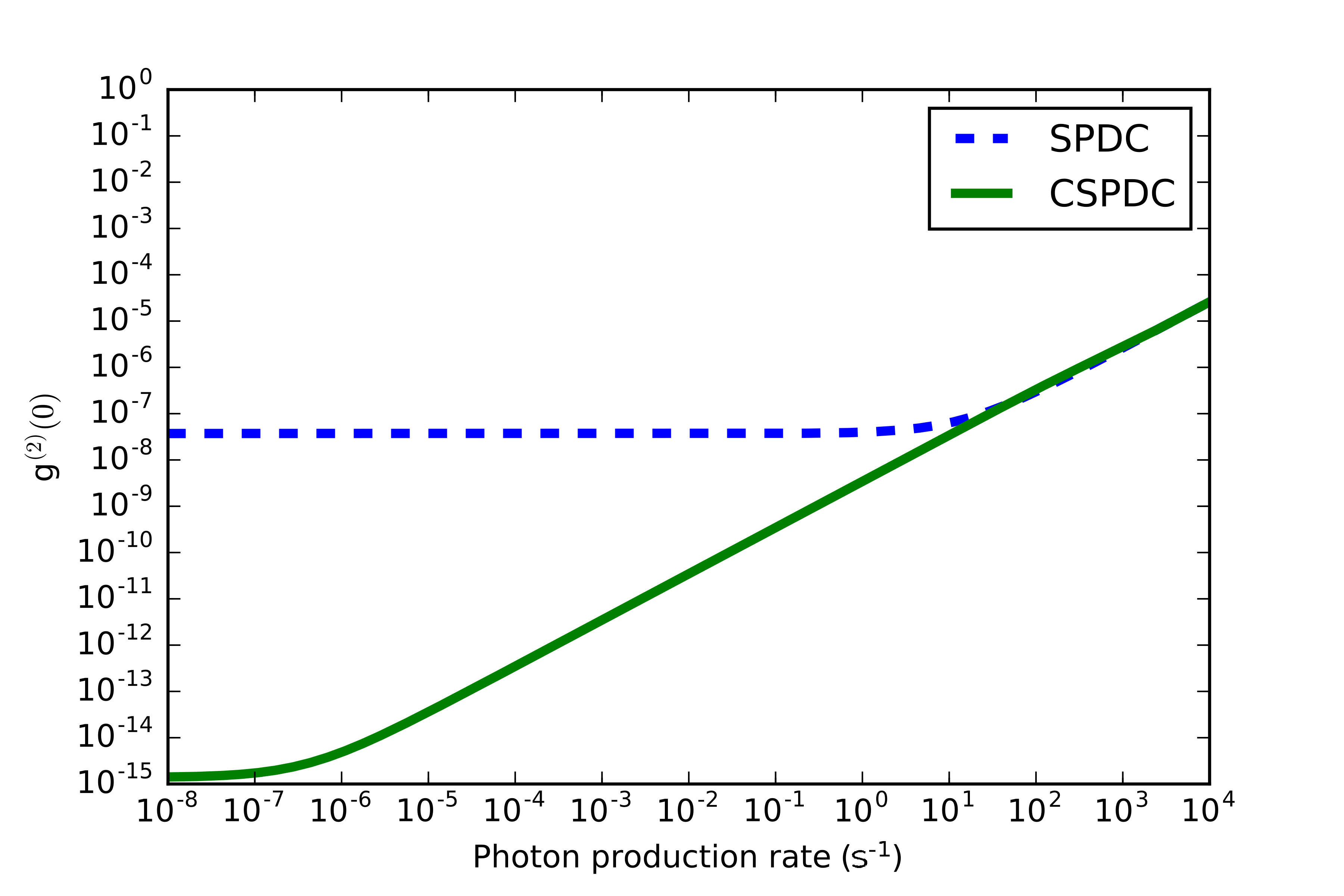}
\captionof{figure}{\label{fig:perf_dets_graph}With ideal $g^{(2)}$ detectors, we find a substantial improvement of the minimal $g^{(2)}$. In this example, $\eta_1 = \eta_2 = 0.7$, $d_1=d_2=$~\SI{10}{\per\second}, $W=$~\SI{2}{\nano\second} and $P=10^{-6}$. }
\end{figure}

The results of the simulation are shown in figure \ref{fig:perf_dets_graph} for a typical SPDC crystal and detector performance. We notice that in this case, where the performance of the $g^{(2)}$ detectors is ignored, CSPDC provides a dramatic improvement over SPDC. While the maximal advantage over SPDC is only achieved at impractically low rates, at 0.01 cps the heralded $g^{(2)}$ is already lowered by two orders of magnitude. 

\subsection{Modeling the $g^{(2)}$ with identical detectors}
In practice, the $g^{(2)}$ detectors will not be perfect, which will significantly impact the measured heralded $g^{(2)}$. Therefore, we now consider a more realistic approach where all the detectors have similar performance. Here $H_{AB} = H_1 = H_2 = H$, and  equation \ref{eq:criterion} becomes

\begin{equation}\label{eq:crit_H_same}
P > \frac{1}{H F(\eta_1)},
\end{equation}

\noindent where 
\begin{equation}
F(\eta_1)=\sqrt{2-\eta_1} + \frac{2-\eta_1}{4}.
\end{equation}

Finally the improvement becomes 
\begin{equation}\label{eq:improvementratio}
\frac{g^{(2)}_S}{g^{(2)}_C} = \frac{1+F(\eta)}{1+\frac{1}{PH}}.
\end{equation}

\begin{figure}[h!]
\begin{center}
    \includegraphics[width=3.5in]{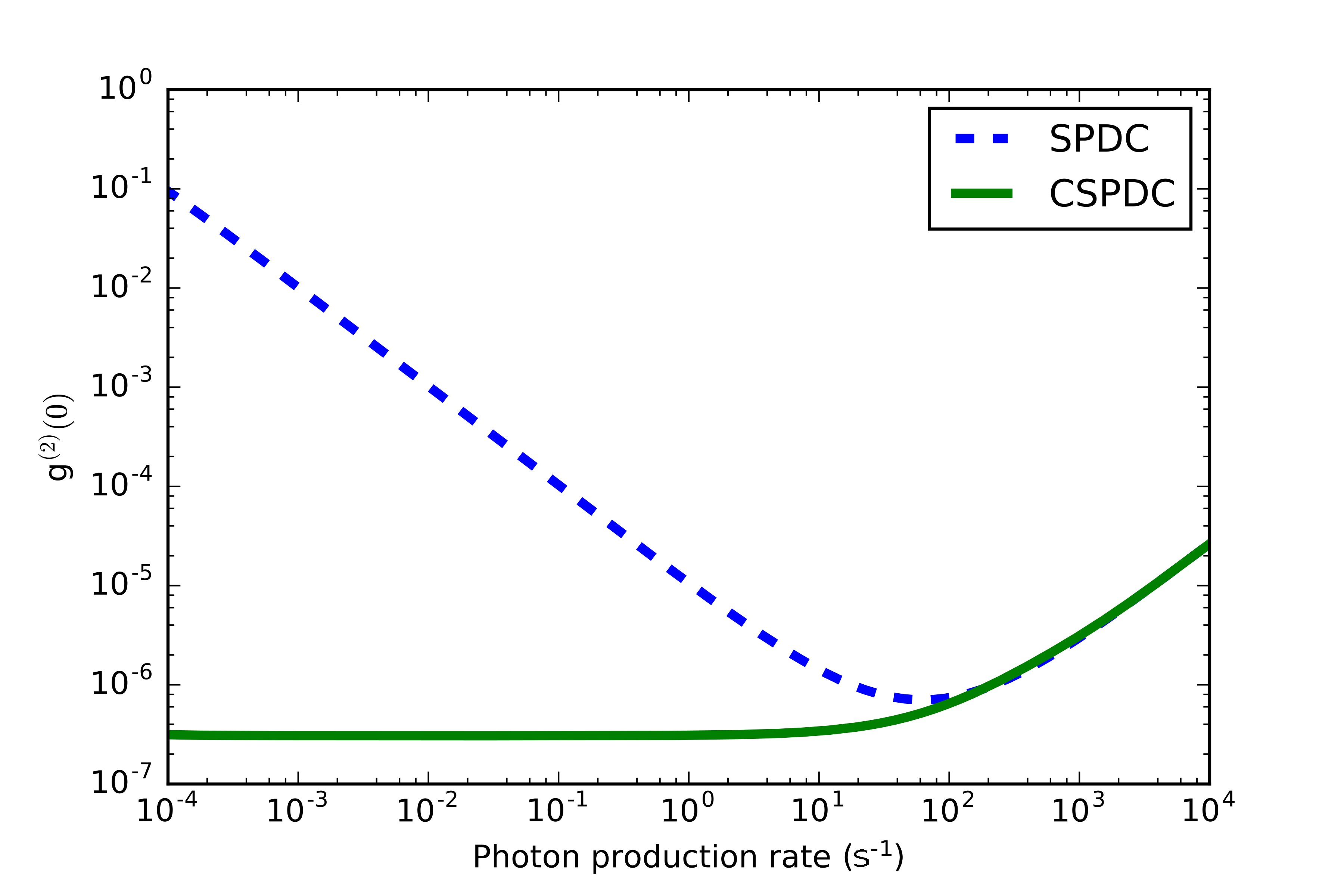}
    \captionof{figure}[Heralded measurement with cascaded SPDC source]{\label{fig:g2} A typical example of the $g^{(2)}$ as a function of pump rate assuming $\eta=0.7$, $P=10^{-6}$, $W=$~\SI{5}{\nano\second} and $d=$~\SI{20}{\per\second} for SPDC and CSPDC. CSPDC maintains the minimal $g^{(2)}$ on an extended range of heralded single photon production rates.}
\end{center}
\end{figure}

Considering that $0 < \eta <1$ we have $1.25 < F(\eta) < 1.91$, which gives a maximal improvement factor of $2.91$ for the minimum $g^{(2)}$ achievable. This is confirmed by the simulation results, shown in Figure  \ref{fig:g2}.
While the improvement in the optimal $g^{(2)}$ is not as substantial as the last case considered, it can be reached while maintaining practical pair production rates. Significantly, CSPDC is advantageous for equal photon production rates. Furthermore, the minimal value of the heralded $g^{(2)}$ is achieved on a broad range of photon production rates. A CSPDC source can therefore remain at the optimal $g^{(2)}$ despite orders of magnitude of change in pump power.

For a given set of detectors, the $g^{(2)}$ is minimized by reducing the pump power to control the pair production rate and is optimized when the pair production rate is low enough that effects from dark counts becomes dominant. However, given that CSPDC decreases the false heralding rates by requiring two heralds, the minimal value of $g^{(2)}$ achievable is lower in cases where equation \ref{eq:crit_H_same} is fulfilled. As shown in figure \ref{fig:g2}, for typical parameters, this is achievable with commercially available detectors.  For example, a shallow junction silicon single-photon avalanche diode (SPAD) detector has a figure of merit above $10^8$ and superconducting nanowire detectors can achieve figures of merit about $10^9$\cite{Eisaman2011}. Materials such as periodically poled lithium niobate (PPLN) in optical waveguides can reach conversion efficiencies of $10^{-6}$\cite{Hubel2010,Tanzilli2001}. Using these elements, CSPDC would be advantageous regardless of Klyshko efficiency.

\section{Conclusion}

Our results confirm that CSPDC allows for a lower heralded $g^{(2)}$ than normal SPDC. This maximal improvement can be achieved using typical detectors and is more stable to fluctuations in pump beam intensity. This yields a process which would be useful in cases where $g^{(2)}$ purity is paramount, for example in linear optical quantum computing. 

In cases where CSPDC is advantageous, the optimal range of the photon production rate is also larger, spanning multiple orders of magnitude, meaning that the value of the $g^{(2)}$ is resistant to even large fluctuations in pump beam intensity. This characteristic is exemplified in figure \ref{fig:g2}.

An important consideration for heralded single photons is the spectral purity of photons, which would be limited by the intrinsic spectral correlations of cascaded downconversion experiments performed to date\cite{Shalm2013}. However, some schemes are able to control these correlations. For example, we expect that that this challenge could be addressed by using recently published schemes such as spectral multiplexing with feed-forward control \cite{Puigibert2017}, by using the SPDC in a type-II waveguide where the crystal's length is controlled\cite{Harder2013} or by manipulating tripartite frequency correlation of the produced triplets pre-heralding\cite{Zhang2018}.

The method used in the simulations can be extended further, by considering, for example, different arrangement of detectors using CSPDC. We also believe that the method extends to more complex heralded single photon source schemes, and for schemes using single-photons as pumps for nonlinear processes.

\section*{Funding}
This research was undertaken, in part, thanks to funding from the Canada Research Chairs program, the Canada Foundation for Innovation, the Natural Sciences and Engineering Research Council of Canada and the New Brunswick Innovation Foundation.

\bibliographystyle{osajnl}

\begin{thebibliography}{10}
\newcommand{\enquote}[1]{``#1''}

\bibitem{Eisaman2011}
M.~D. Eisaman, J.~Fan, A.~Migdall, and S.~V. Polyakov, \enquote{Invited review
  article: Single-photon sources and detectors,} Rev. Sci. Instrum.
 \textbf{82}, 071101 (2011).

\bibitem{Schiavon2016}
M.~Schiavon, G.~Vallone, F.~Ticozzi, and P.~Villoresi, \enquote{Heralded
  single-photon sources for quantum-key-distribution applications,} Phys. Rev. A \textbf{93}, 012331 (2016).

\bibitem{Migdall2002a}
A.~L. Migdall, D.~Branning, and S.~Castelletto, \enquote{Tailoring
  single-photon and multiphoton probabilities of a single-photon on-demand
  source,} Phys. Rev. A \textbf{66} (2002).

\bibitem{Grice2001}
W.~P. Grice, A.~U'Ren, and I.~A. Walmsley, \enquote{Eliminating frequency and
  space-time correlations in multiphoton states,} Phys. Rev. A \textbf{64},
  063815 (2001).

\bibitem{URen2005}
A.~B. U’Ren, C.~Silberhorn, J.~L. Ball, K.~Banaszek, and I.~A. Walmsley,
  \enquote{Characterization of the nonclassical nature of conditionally
  prepared single photons,} Phys. Rev. A \textbf{72}, 021802 (2005).

\bibitem{Mosley2008a}
P.~J. Mosley, J.~S. Lundeen, B.~J. Smith, P.~Wasylczyk, A.~B. U’Ren,
  C.~Silberhorn, and I.~A. Walmsley, \enquote{Heralded generation of ultrafast
  single photons in pure quantum states,} Phys. Rev. Lett. \textbf{100},
  133601 (2008).

\bibitem{Mosley2008}
P.~J. Mosley, J.~S. Lundeen, B.~J. Smith, and I.~A. Walmsley,
  \enquote{Conditional preparation of single photons using parametric
  downconversion: a recipe for purity,} New J. Phys. \textbf{10},
  093011 (2008).

\bibitem{Levine2010}
Z.~H. Levine, J.~Fan, J.~Chen, A.~Ling, and A.~Migdall, \enquote{Heralded,
  pure-state single-photon source based on a potassium titanyl phosphate
  waveguide,} Opt. Express \textbf{18}, 3708 (2010).

\bibitem{Jeffrey2004}
E.~Jeffrey, N.~A. Peters, and P.~G. Kwiat, \enquote{Towards a periodic
  deterministic source of arbitrary single-photon states,} New J. Phys. \textbf{6}, 100--100 (2004).

\bibitem{Shapiro2007}
J.~H. Shapiro and F.~N. Wong, \enquote{On-demand single-photon generation using
  a modular array of parametric downconverters with electro-optic polarization
  controls,} Opt. Lett. \textbf{32}, 2698 (2007).

\bibitem{Ma2011}
X.~S. Ma, S.~Zotter, J.~Kofler, T.~Jennewein, and A.~Zeilinger,
  \enquote{Experimental generation of single photons via active multiplexing,}
  Phys. Rev. A \textbf{83} (2011).

\bibitem{Collins2013}
M.~Collins, C.~Xiong, I.~Rey, T.~Vo, J.~He, S.~Shahnia, C.~Reardon, T.~Krauss,
  M.~Steel, A.~Clark, and B.~Eggleton, \enquote{Integrated spatial multiplexing
  of heralded single-photon sources,} Nat. Commun. \textbf{4} (2013).

\bibitem{Xiong2016}
C.~Xiong, X.~Zhang, Z.~Liu, M.~Collins, A.~Mahendra, L.~Helt, M.~Steel, D.-Y.
  Choi, C.~Chae, P.~Leong \emph{et~al.}, \enquote{Active temporal multiplexing
  of indistinguishable heralded single photons,} Nat. Commun.
  \textbf{7} (2016).

\bibitem{Mendoza2016}
G.~J. Mendoza, R.~Santagati, J.~Munns, E.~Hemsley, M.~Piekarek,
  E.~Mart{\'{\i}}n-L{\'{o}}pez, G.~D. Marshall, D.~Bonneau, M.~G. Thompson, and
  J.~L. O'Brien, \enquote{Active temporal and spatial multiplexing of photons,}
  Optica \textbf{3}, 127 (2016).

\bibitem{Puigibert2017}
M.~Grimau~Puigibert, G.~H. Aguilar, Q.~Zhou, F.~Marsili, M.~D. Shaw, V.~B.
  Verma, S.~W. Nam, D.~Oblak, and W.~Tittel, \enquote{Heralded single photons
  based on spectral multiplexing and feed-forward control,} Phys. Rev. Lett.
  \textbf{119}, 083601 (2017).

\bibitem{Kaneda2015}
F.~Kaneda, B.~G. Christensen, J.~J. Wong, H.~S. Park, K.~T. McCusker, and P.~G.
  Kwiat, \enquote{Time-multiplexed heralded single-photon source,} Optica
  \textbf{2}, 1010--1013 (2015).

\bibitem{Cabello2012}
A.~Cabello and F.~Sciarrino, \enquote{Loophole-free bell test based on local
  precertification of photon's presence,} Phys. Rev. X \textbf{2}, 021010
  (2012).

\bibitem{Hubel2010}
H.~{H{\"u}bel}, D.~R. {Hamel}, A.~{Fedrizzi}, S.~{Ramelow}, K.~J. {Resch}, and
  T.~{Jennewein}, \enquote{{Direct generation of photon triplets using cascaded
  photon-pair sources},} Nature \textbf{466}, 601--603 (2010).

\bibitem{Krapick2016}
S.~Krapick, B.~Brecht, H.~Herrmann, V.~Quiring, and C.~Silberhorn,
  \enquote{On-chip generation of photon-triplet states,} Opt. Express
  \textbf{24}, 2836--2849 (2016).

\bibitem{Ding2015}
D.-S. Ding, W.~Zhang, S.~Shi, Z.-Y. Zhou, Y.~Li, B.-S. Shi, and G.-C. Guo,
  \enquote{Hybrid-cascaded generation of tripartite telecom photons using an
  atomic ensemble and a nonlinear waveguide,} Optica \textbf{2}, 642--645
  (2015).

\bibitem{Meyer-Scott2016}
E.~Meyer-Scott, D.~McCloskey, K.~Go\l{}os, J.~Z. Salvail, K.~A.~G. Fisher,
  D.~R. Hamel, A.~Cabello, K.~J. Resch, and T.~Jennewein, \enquote{Certifying
  the presence of a photonic qubit by splitting it in two,} Phys. Rev. Lett.
  \textbf{116}, 070501 (2016).

\bibitem{Grangier1986}
P.~Grangier, G.~Roger, and A.~Aspect, \enquote{Experimental evidence for a
  photon anticorrelation effect on a beam splitter: A new light on
  single-photon interferences,} Europhys. Lett. \textbf{1}, 173
  (1986).

\bibitem{Brassard2000}
G.~Brassard, N.~Lütkenhaus, T.~Mor, and B.~C. Sanders, \enquote{Limitations on
  {Practical} {Quantum} {Cryptography},} Phys. Rev. Lett. \textbf{85},
  1330--1333 (2000).

\bibitem{Jennewein2011}
T.~Jennewein, M.~Barbieri, and A.~G. White, \enquote{Single-photon device
  requirements for operating linear optics quantum computing outside the
  post-selection basis,} J. Mod. Optic. \textbf{58}, 276--287 (2011).

\bibitem{Senellart2017}
P.~Senellart, G.~Solomon, and A.~G. White, \enquote{High-performance
  semiconductor quantum-dot single-photon sources,} Nat. Nanotechnol.
  \textbf{12}, 1026--1039 (2017).

\bibitem{Bussieres2008}
F.~Bussi{\`{e}}res, J.~A. Slater, N.~Godbout, and W.~Tittel, \enquote{Fast and
  simple characterization of a photon pair source,} Opt. Express \textbf{16},
  17060 (2008).

\bibitem{Klyshko1980}
D.~N. Klyshko, \enquote{Use of two-photon light for absolute calibration of
  photoelectric detectors,} Sov. J. Quantum Electron. \textbf{10},
  1112 (1980).

\bibitem{Hadfield2009}
R.~H. Hadfield, \enquote{Single-photon detectors for optical quantum
  information applications,} Nat. Photonics \textbf{3}, 696--705 (2009).

\bibitem{Takesue2010}
H.~Takesue and K.~Shimizu, \enquote{Effects of multiple pairs on visibility
  measurements of entangled photons generated by spontaneous parametric
  processes,} Opt. Commun. \textbf{283}, 276--287 (2010).

\bibitem{Shalm2013}
L.~K. Shalm, D.~R. Hamel, Z.~Yan, C.~Simon, K.~J. Resch, and T.~Jennewein,
  \enquote{Three-photon energy-time entanglement,} Nat. Phys. \textbf{9},
  19--22 (2013).

\bibitem{Zhang2018}
Q.-Y. Zhang, G.-T. Xue, P.~Xu, Y.-X. Gong, Z.~Xie, and S.~Zhu,
  \enquote{Manipulation of tripartite frequency correlation under extended
  phase matchings,} Phys. Rev. A \textbf{97} (2018).

\bibitem{Tanzilli2001}
S.~Tanzilli, H.~De~Riedmatten, W.~Tittel, H.~Zbinden, P.~Baldi, M.~De~Micheli,
  D.~Ostrowsky, and N.~Gisin, \enquote{Highly efficient photon-pair source
  using periodically poled lithium niobate waveguide,} Electron. Lett.
  \textbf{37}, 26 (2001).

\bibitem{Harder2013}
G.~Harder, V.~Ansari, B.~Brecht, T.~Dirmeier, C.~Marquardt, and C.~Silberhorn,
  \enquote{An optimized photon pair source for quantum circuits,} Opt. Express
  \textbf{21}, 13975--13985 (2013).

\end{thebibliography}

\end{document}